\documentstyle[preprint,epsfig,aps,floats,tighten]{revtex}

\def\ppbar{$p\overline{p} $}            
\def\qqbar{$q\overline{q}$}             
            \def\ttbar{$t\overline{t}$}             
\def\pt{$p_T$}                          
\def\et{$E_T$}                          
\def\met{\mbox{${\hbox{$E$\kern-0.6em\lower-.1ex\hbox{/}}}_T$}} 
\def\ipb{pb$^{-1}$}                     
\def\gevcc{GeV}                         
\def\gevc{GeV}                          
\def\gev{GeV}                           
\def\D0{D\O}                            
\def\etal{{\sl et al.}}                   
\def\d0draft{}

\newcommand{\LEGA}{{\sc legacy}}

\newcommand{\PYTH}{{\sc pythia}}

\newcommand{\HERW}{{\sc herwig}}

\def\wb{$W$}
\def\zb{$Z$}

\def\tev{TeV}
\def\Dzero{D\O}
\def\dsdpt{$d\sigma/dp_T$}

\def\gevsq{GeV$^2$}

\def\zbee{$Z\rightarrow e^+e^-$}

\def\ttbar{{$t\overline{t}$}}
\def\zbtt{{$Z\rightarrow \tau^+\tau^-$}}

\begin{document}

\onecolumn
\title{ Differential Production Cross Section of \zb\ Bosons
        as a Function of Transverse
        Momentum at $\sqrt{s}=1.8$ \tev  }

%
\author{
B.~Abbott,$^{45}$
M.~Abolins,$^{42}$
V.~Abramov,$^{18}$
B.S.~Acharya,$^{11}$
I.~Adam,$^{44}$
D.L.~Adams,$^{54}$
M.~Adams,$^{28}$
S.~Ahn,$^{27}$
V.~Akimov,$^{16}$
G.A.~Alves,$^{2}$
% ...
\\
\vskip 0.30cm
\centerline{(D\O\ Collaboration)}
\vskip 0.30cm
}
\address{
\centerline{$^{1}$Universidad de Buenos Aires, Buenos Aires, Argentina}
\centerline{$^{2}$LAFEX, Centro Brasileiro de Pesquisas F{\'\i}sicas,
                  Rio de Janeiro, Brazil}
\centerline{$^{3}$Universidade do Estado do Rio de Janeiro,
                  Rio de Janeiro, Brazil}
\centerline{$^{4}$Institute of High Energy Physics, Beijing,
                  People's Republic of China}
\centerline{$^{5}$Universidad de los Andes, Bogot\'{a}, Colombia}
\centerline{$^{6}$Universidad San Francisco de Quito, Quito, Ecuador}
\centerline{$^{7}$Institut des Sciences Nucl\'eaires, IN2P3-CNRS,
                  Universite de Grenoble 1, Grenoble, France}
\centerline{$^{8}$DAPNIA/Service de Physique des Particules, CEA, Saclay,
                  France}
\centerline{$^{9}$Panjab University, Chandigarh, India}
\centerline{$^{10}$Delhi University, Delhi, India}
\centerline{$^{11}$Tata Institute of Fundamental Research, Mumbai, India}
\centerline{$^{12}$Kyungsung University, Pusan, Korea}
\centerline{$^{13}$Seoul National University, Seoul, Korea}
\centerline{$^{14}$CINVESTAV, Mexico City, Mexico}
\centerline{$^{15}$Institute of Nuclear Physics, Krak\'ow, Poland}
\centerline{$^{16}$Institute for Theoretical and Experimental Physics,
                   Moscow, Russia}
\centerline{$^{17}$Moscow State University, Moscow, Russia}
\centerline{$^{18}$Institute for High Energy Physics, Protvino, Russia}
\centerline{$^{19}$Lancaster University, Lancaster, United Kingdom}
\centerline{$^{20}$University of Arizona, Tucson, Arizona 85721}
\centerline{$^{21}$Lawrence Berkeley National Laboratory and University of
                   California, Berkeley, California 94720}
\centerline{$^{22}$University of California, Davis, California 95616}
\centerline{$^{23}$University of California, Irvine, California 92697}
\centerline{$^{24}$University of California, Riverside, California 92521}
\centerline{$^{25}$Florida State University, Tallahassee, Florida 32306}
\centerline{$^{26}$University of Hawaii, Honolulu, Hawaii 96822}
\centerline{$^{27}$Fermi National Accelerator Laboratory, Batavia,
                   Illinois 60510}
\centerline{$^{28}$University of Illinois at Chicago, Chicago,
                   Illinois 60607}
\centerline{$^{29}$Northern Illinois University, DeKalb, Illinois 60115}
\centerline{$^{30}$Northwestern University, Evanston, Illinois 60208}
\centerline{$^{31}$Indiana University, Bloomington, Indiana 47405}
\centerline{$^{32}$University of Notre Dame, Notre Dame, Indiana 46556}
\centerline{$^{33}$Purdue University, West Lafayette, Indiana 47907}
\centerline{$^{34}$Iowa State University, Ames, Iowa 50011}
\centerline{$^{35}$University of Kansas, Lawrence, Kansas 66045}
\centerline{$^{36}$Kansas State University, Manhattan, Kansas 66506}
\centerline{$^{37}$Louisiana Tech University, Ruston, Louisiana 71272}
\centerline{$^{38}$University of Maryland, College Park, Maryland 20742}
\centerline{$^{39}$Boston University, Boston, Massachusetts 02215}
\centerline{$^{40}$Northeastern University, Boston, Massachusetts 02115}
\centerline{$^{41}$University of Michigan, Ann Arbor, Michigan 48109}
\centerline{$^{42}$Michigan State University, East Lansing, Michigan 48824}
\centerline{$^{43}$University of Nebraska, Lincoln, Nebraska 68588}
\centerline{$^{44}$Columbia University, New York, New York 10027}
\centerline{$^{45}$New York University, New York, New York 10003}
\centerline{$^{46}$University of Rochester, Rochester, New York 14627}
\centerline{$^{47}$State University of New York, Stony Brook,
                   New York 11794}
\centerline{$^{48}$Brookhaven National Laboratory, Upton, New York 11973}
\centerline{$^{49}$Langston University, Langston, Oklahoma 73050}
\centerline{$^{50}$University of Oklahoma, Norman, Oklahoma 73019}
\centerline{$^{51}$Brown University, Providence, Rhode Island 02912}
\centerline{$^{52}$University of Texas, Arlington, Texas 76019}
\centerline{$^{53}$Texas A\&M University, College Station, Texas 77843}
\centerline{$^{54}$Rice University, Houston, Texas 77005}
}

\maketitle

\begin{abstract}
We present a measurement of the transverse momentum
distribution of \zb\ bosons produced in \ppbar\ collisions at
$\sqrt{s}=1.8$ \tev\ using data collected by the \Dzero\ experiment
at the Fermilab Tevatron Collider during 1994--1996.  We find good
agreement between our data and a current resummation calculation. We
also use our data to extract values of the non-perturbative parameters
for a particular version of the resummation formalism, obtaining
significantly more precise values than previous determinations.

\end{abstract}


\clearpage
We report a new measurement\cite{PRD,Casey_thesis}
of the differential cross section with respect to transverse
momentum (\dsdpt) of the \zb\ boson in the dielectron channel
with statistics and precision greatly improved beyond previous
measurements~\cite{UA2_ptz,CDF_ptz}. The measurement of \dsdpt\
of the \zb\ boson provides a sensitive test of QCD at high-$Q^2$.
At small transverse momentum (\pt), where the cross section is highest,
uncertainties in the phenomenology of vector boson production
have contributed significantly to the uncertainty in the mass of
the \wb\ boson.  Due to its similar production characteristics
and the fact that the decay electrons can be very well-measured,
the \zb\ provides a good laboratory for evaluating the
phenomenology of vector boson production.

In the parton model, \zb\ bosons are produced in
collisions of \qqbar\ constituents of the proton and
antiproton. The fact that observed \zb\ bosons have finite
\pt\ can be attributed
to gluon radiation from the colliding partons prior to their
annihilation. In standard perturbative QCD (pQCD), the
cross section for \zb\ boson production is calculated by expanding
in powers of the strong coupling
constant, $\alpha_s$. This procedure works well when $p_T^2\sim
Q^2$ with $Q=M_Z$. However, when $p_T\ll Q$, correction terms that are
proportional to $\alpha_s\ln(Q^2/{p_T^2})$ become significant, and
the cross section diverges at small \pt. This difficulty is surmounted
by reordering the perturbative series
through a technique called {\it resummation}
\cite{CSS,DaviesStirling,AEGM,DWS,ArnoldReno,AK,LadinskyYuan,BalazsYuan,EllisVeseli}.
Although this technique extends the applicability of pQCD to lower
values of $p_T$, a more fundamental barrier is encountered when
\pt\ approaches $ \Lambda_{\text{QCD}}$. In this region,
$\alpha_s$ becomes large and the perturbative calculation is no
longer valid. In order to account for the non-perturbative contribution,
a phenomenological form factor must be
invoked, which contains several parameters that must be tuned to data
~\cite{DWS,AK,LadinskyYuan}.

The resummation may be carried
out in impact-parameter ($b$) space via a Fourier transform, or in
transverse momentum space. Both formalisms require a
non-perturbative function to describe the low-\pt\ region beyond some
cut-off value $b_{max}$ or $p_{Tlim}$ and they merge to
the fixed-order perturbation theory at $p_T\sim Q$. The current
state-of-the-art for
the $b$-space formalism resums terms to next-to-next-to-next-to-leading-log
and includes fixed-order
terms to ${\cal{O}}({\alpha_s^2})$~\cite{LadinskyYuan}. Similarly,
the $p_T$-space formalism resums terms to next-to-next-to-leading-log
 and includes fixed-order
terms to ${\cal{O}}({\alpha_s})$~\cite{EllisVeseli}.

In the $b$-space formalism, the resummed cross section is modified at
large $b$ (above $b_{max}$) by $\exp(-S_{\rm{NP}}(b,Q^2))$. The
form factor $S_{\rm{NP}}(b,Q^2)$ has a general renormalization group
invariant form,
but requires a specific choice of parameterization when making predictions.
A possible choice, suggested by Ladinsky and Yuan~\cite{LadinskyYuan}, is
\begin{eqnarray}
  S_{\rm NP}(b,Q^2)= \nonumber \hspace{-1.5cm} \\
  & g_1b^2+g_2b^2\ln({Q^2\over Q_o^2})+
                      g_1g_3b\ln(100x_ix_j),
\end{eqnarray}
where $x_i$ and $x_j$ are the fractions of incident hadron momenta
carried by the colliding partons and $g_i$ are the non-perturbative
parameters. An earlier parameterization by
Davies, Webber, and Stirling~\cite{DWS} corresponds to the
above with $g_3\equiv 0$. For measurements at the Fermilab Tevatron
at $Q^2=M_Z^2$, the calculation is most sensitive to the value of $g_2$
and quite insensitive to the value of $g_3$.

In the $p_T$-space formalism, the resummed cross section is modified at
low-\pt\ (below $p_{Tlim}$) by multiplying the cross section by
$F_{\rm{NP}}(p_T)$.
In this case, the form of the non-perturbative function is not constrained by
renormalization group invariance. The choice suggested by Ellis and
Veseli~\cite{EllisVeseli}, is
\begin{equation}
       \tilde{F}_{\rm NP}(p_T)=1-e^{-\tilde{a}p^2_T}
\end{equation}
where $\tilde{a}$ is a non-perturbative parameter.

Previously published measurements of the differential cross section for
$Z$ boson production have been limited primarily by statistics
(candidate samples of a few hundred events).
This measurement is based on a sample of 6407 \zbee\ events,
corresponding to an integrated luminosity of $\approx 111$ \ipb, collected
with the \Dzero\ detector~\cite{D0Nim} in 1994-1996. A recent measurement
by the CDF Collaboration has a similar number of events~\cite{cdf_prl}.

Electrons are detected in the uranium/liquid-argon calorimeter
    with a fractional energy resolution of $\approx 15\%/\sqrt{E(\rm GeV)}$.
    The calorimeter has a transverse granularity at the electron
    shower maximum of $\Delta\eta \times
    \Delta\phi = 0.05 \times 0.05$, where $\eta$ is the pseudorapidity
    and $\phi$ is the azimuthal angle.
 The two
electron candidates in the event with the highest transverse energy (\et),
both having \et $>$ 25 \gev, are
used to reconstruct the \zb\ boson candidate. One electron is required
to be in the central region, $|\eta_{\rm det}|<1.1$, and the second
electron may be either in the central or in the forward region,
$1.5<|\eta_{\rm det}|<2.5$, where $\eta_{\rm det}$ refers to the value
of $\eta$ obtained by assuming that the shower originates from the center
of the detector. Offline, both
electrons are required to be isolated and to satisfy
cluster-shape requirements. Additionally, at least one of the
electrons is required to have a matching track in the drift
chamber system that points to
the reconstructed calorimeter cluster.

Both the acceptance and the theory predictions modified by the
\Dzero\ detector resolution are
calculated using a simulation technique originally developed for
measuring the mass of the \wb\ boson~\cite{D0Wmass},
with minor modifications required by changes in selection
criteria.
The four-momentum of the \zb\ boson is obtained by generating the
mass of the \zb\ according to an energy-dependent Breit-Wigner
lineshape. The \pt\ and rapidity of the \zb\ boson are chosen
randomly from two-dimensional grids created using the computer program
\LEGA~\cite{BalazsYuan}, which calculates the \zb\ boson cross section
for a given \pt, rapidity, and mass of the \zb\ boson.
The positions and energies of the electrons are smeared
according to the measured resolutions, and corrected for offsets in
energy scale caused by the underlying event and recoil particles
that overlap the calorimeter towers. Underlying events are modeled
using data from random inelastic \ppbar\ collisions of the same
luminosity profile as the \zb\ boson sample.
The electron energy and angular resolutions are tuned to reproduce
the observed width of the mass distribution at the
\zb -boson resonance and the difference between the
reconstructed vertex positions of the electrons.

We determine the shape of the efficiency of the event selection criteria
as a function of \pt\ using \zbee\  events generated with
\HERW \cite{HERWIG}, smeared with the \Dzero\ detector resolutions,
and overlaid on randomly selected zero bias \ppbar\ collisions. This
simulation models
the effects of the underlying event and jet activity on the selection
of the electrons. The absolute
efficiency is obtained from \zbee\ data \cite{wzcross}. The values
of the efficiency times acceptance range from 26-37\% for \pt\ below
200 \gev\ and is 53\% for \pt\ above 200 \gev.

The primary background arises
from multiple-jet production from QCD processes in which two jets
pass the electron selection criteria. We use several
\Dzero\ data sets for estimating this background---direct-$\gamma$ events,
dijet events,
and dielectron events in which both electrons fail quality
criteria---all of which have very similar kinematic characteristics~\cite{PRD}.
The  level of the multijet background is determined
by fitting the $ee$ invariant mass in the range $60<M_{ee}<120$ \gevcc\
to a linear combination of Monte Carlo \zbee\ signal events
(using \PYTH~\cite{PYTHIA}) and background (from direct-$\gamma$ events).
We assign a systematic uncertainty to this measurement by
varying the choice of mass window used in the fit, and by changing
the background sample among those mentioned above.
We estimate the total multijet background level to be (4.4$\pm$0.9)\%.
The  direct-$\gamma$ sample is used to
parameterize the shape of the background distribution
as a function of \pt\ .
Backgrounds from other sources,
such as \zbtt, \ttbar, and diboson production,
are negligible.

We use the data corrected for background, acceptance, and efficiency,
to determine the best value of the non-perturbative parameter, $g_2$,
given our data. In the fit, we fix
$g_1$ and $g_3$ to the values obtained in~\cite{LadinskyYuan} and
vary the value of $g_2$. We use the CTEQ4M pdf.
The prediction is smeared with the known detector resolutions,
and the result fitted to our data. The resulting
$\chi^2$ distribution as a function of $g_2$ is well-behaved and
parabolic, yielding a value of $g_2=0.59\pm0.06$ \gevsq, considerably
more precise than previous determinations.
For completeness, we also fit the individual values of $g_1$ and
$g_3$, with the other two parameters fixed to their published
values~\cite{LadinskyYuan}.  We obtain $g_1=0.09\pm 0.03$ \gevsq\ and
$g_3=-1.1\pm0.6$ \gev$^{-1}$. Both results are
consistent with the values of Ref.~\cite{LadinskyYuan}.

\begin{figure}[t]
  \centerline{\psfig{figure=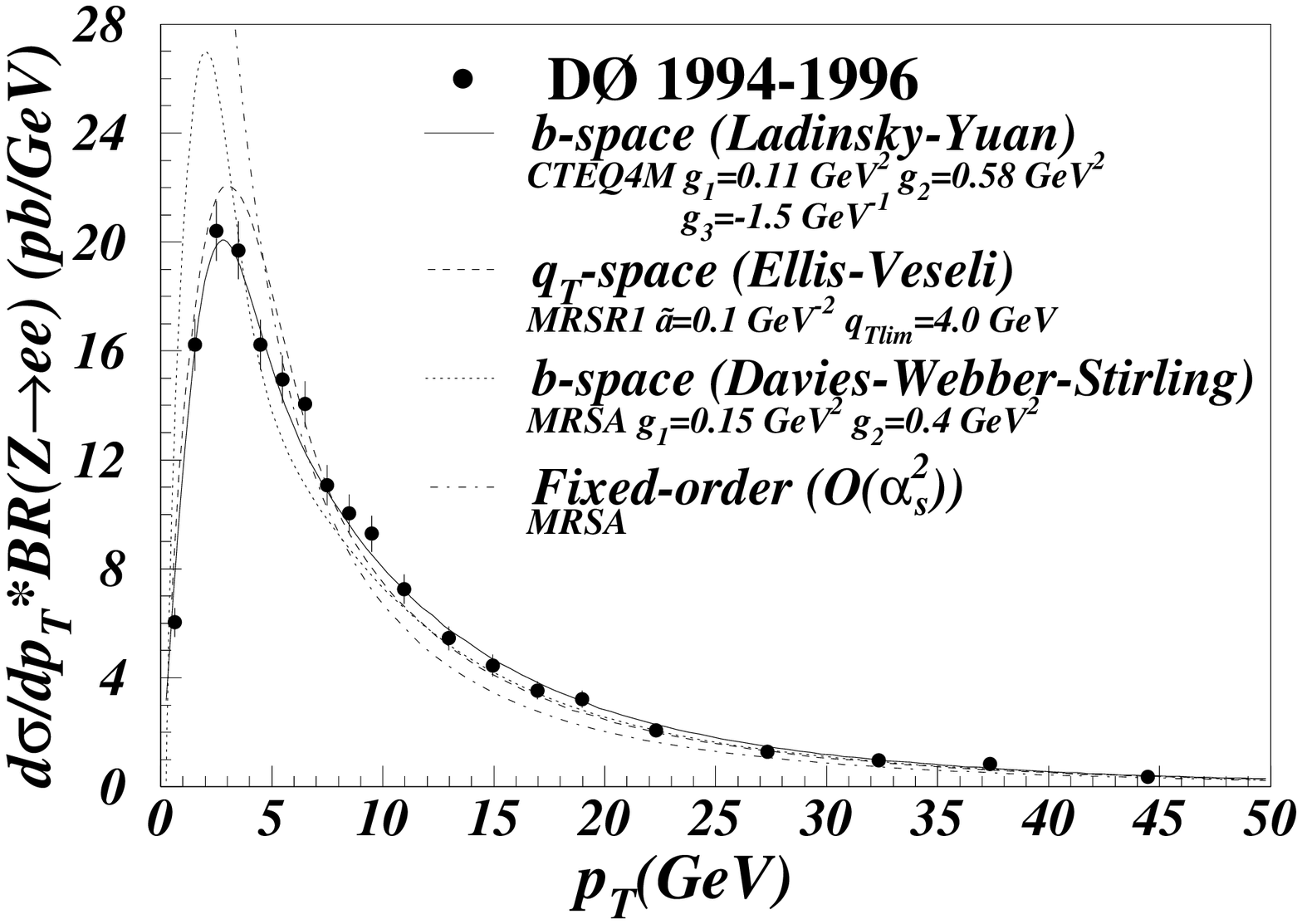,width=3.2in}}
  \vskip -0.7cm
  \centerline{\epsfig{figure=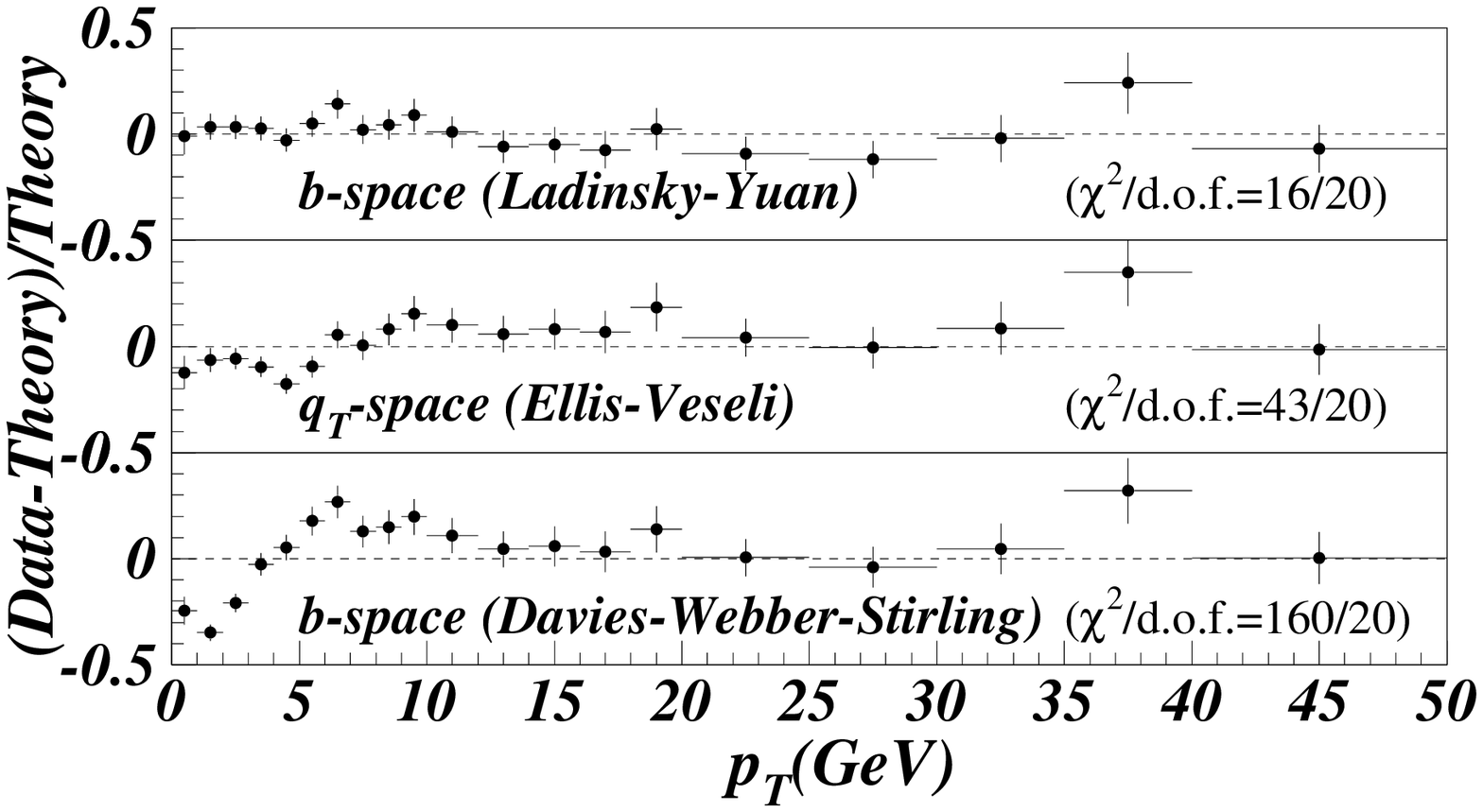,width=3.2in}}
  \caption{The differential cross section as a function of
  \pt\ compared to the resummation calculation with
three different published parameterizations of the non-perturbative
region and to the fixed-order calculation. Also shown are the
fractional differences between the data and each of the resummation
predictions.}
  \label{fig:dsdpt_resum}
\end{figure}

To determine the true \dsdpt, we correct the measured
cross section for effects of detector smearing, using the ratio
of generated to resolution-smeared ansatz \pt\ distributions.
We use the calculation from
\LEGA\ as our ansatz function, with the $g_2$
determined from our fit.
The largest smearing correction occurs at low-\pt, where
smearing causes the largest fractional change in \pt\ and where
the kinematic boundary at \pt$=0$ produces non-Gaussian smearing.
The correction is 18.5\% in the first bin, decreasing to about 2\%
at 5 \gev. For all \pt\ values above 5 \gev, the correction is $\lesssim 5$\%.
Systematic uncertainties arising from the choice of ansatz function
are evaluated by varying $g_2$ within $\pm
1$ standard deviation of the best-fit values. Additional uncertainties are
evaluated by  varying the detector
resolutions by $\pm 1$ standard deviation from the nominal values.
The effect of these variations is negligible relative
to the other uncertainties in the measurement.

\begin{figure}[t]
  \centerline{\psfig{figure=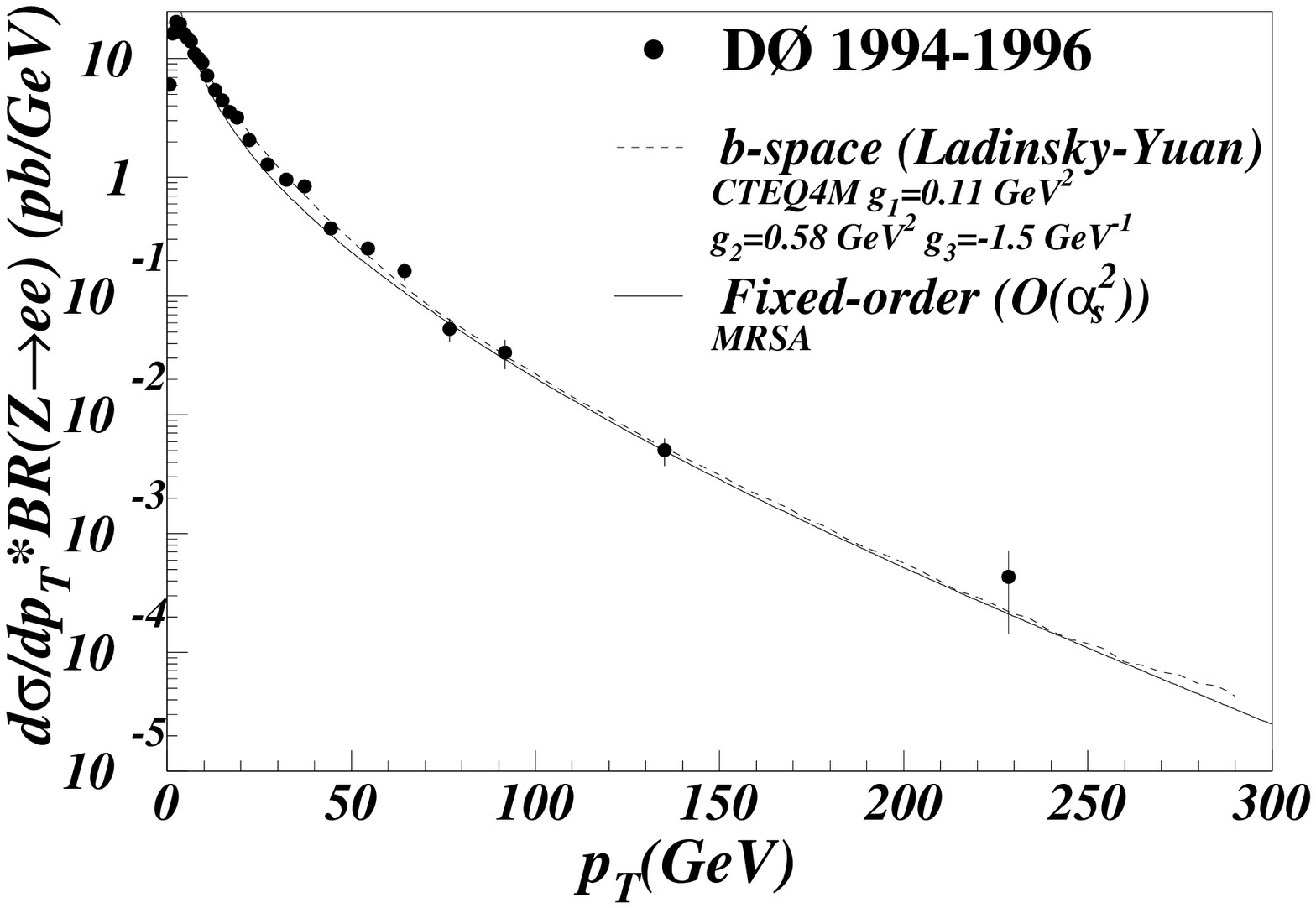,width=3.2in}}
  \vskip -0.7cm
  \centerline{\epsfig{figure=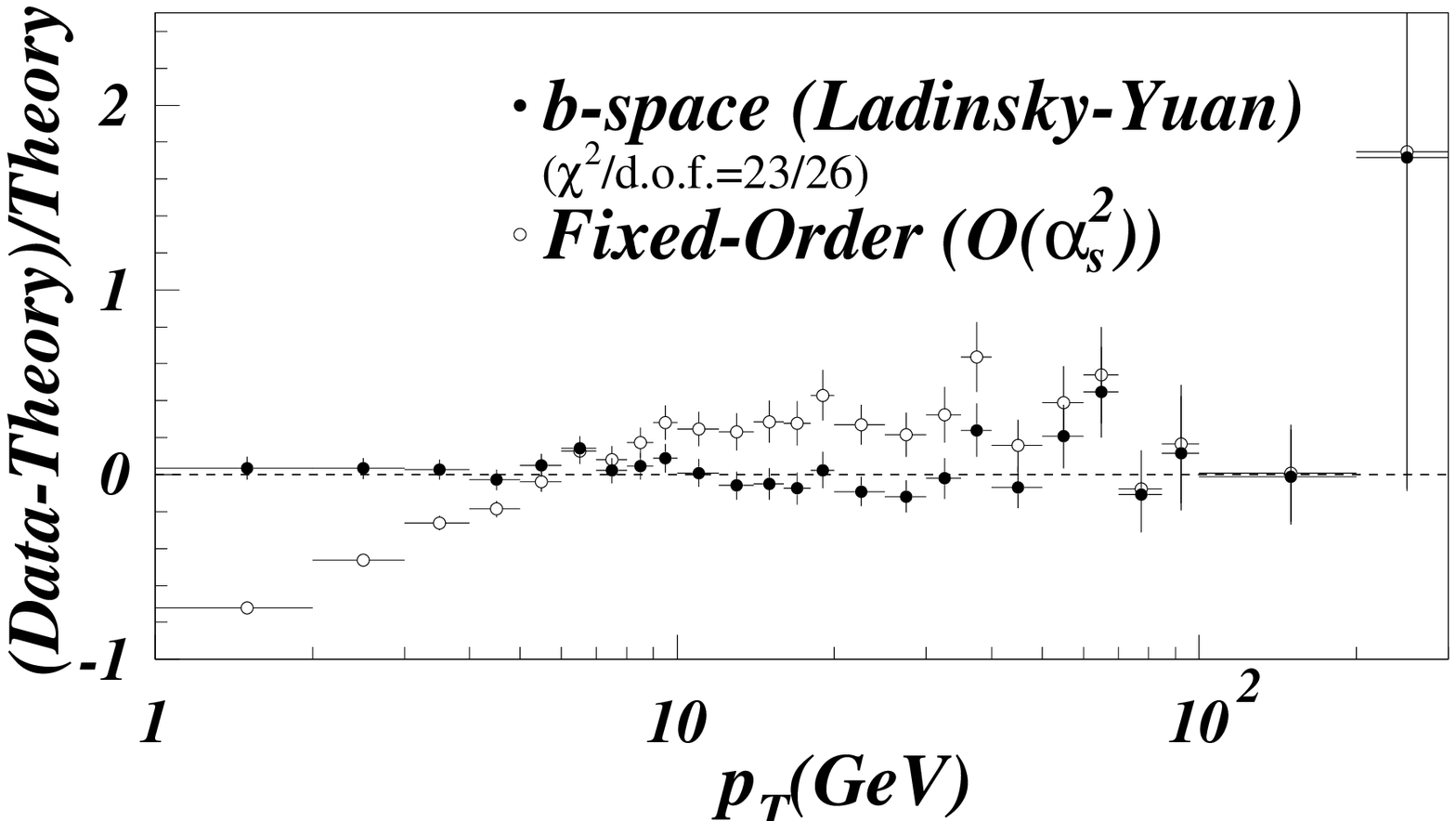,width=3.2in}}
  \caption{A comparison of the data to the resummed and fixed-order
  (${{\cal O}(\alpha^2_s)}$) calculations. Also shown are the
   fractional differences between the data and the resummed and
   fixed-order calculations. The uncertainties shown include both
   statistical and systematic uncertainties (other than an overall
   normalization uncertainty due to the luminosity uncertainty).}
  \label{fig:dsdpt_nlo_resum}
\end{figure}

Table \ref{tab:prl_results} shows the values of
${d\sigma(Z\rightarrow e^+e^-)/dp_T}$.
The uncertainties on the data points include statistical
and systematic contributions.
\noindent An additional normalization
uncertainty of $\pm$4.4\%\ arises from the uncertainty on the integrated
luminosity~\cite{wzcross} that
is not included in any of the plots nor in the table, but must be taken
into account in any fits involving an absolute normalization.

Figure~\ref{fig:dsdpt_resum} shows the
final differential cross section, corrected for the \Dzero\
detector resolutions, compared to the fixed-order calculation and
the resummation calculation with three different parameterizations
of the non-perturbative region using published values of the
non-perturbative parameters. Also shown are the
fractional differences of the data from the considered resummation
predictions. The data are normalized to the
measured \zbee\ cross section (221 pb~\cite{wzcross}) and the predictions
are absolutely normalized.
We observe the best agreement with
the Ladinsky-Yuan parameters for the $b$-space formalism; however,
we expect that fits to the data using the Davies-Weber-Stirling ($b$-space)
or Ellis-Veseli ($p_T$-space) parameterizations of the non-perturbative
functions could describe the data similarly well.

Figure~\ref{fig:dsdpt_nlo_resum} shows the measured differential
cross section compared to the fixed-order calculation and the
resummation calculation using the Ladinsky-Yuan parameterization.
We observe strong disagreement between the data and the fixed-order
prediction in the shape for all but the highest values of \pt. We
attribute this to the divergence of the next-to-leading-order
calculation at \pt$=0$, and a
significant enhancement of the cross section relative to the
prediction at moderate values of \pt. This disagreement confirms
the presence of contributions from soft gluon emission,
which are accounted for in the resummation formalisms.

In summary, we have measured the inclusive differential cross section
of the \zb\ boson as a function
of its transverse momentum. With the enhanced precision of this
measurement over those previous, we can probe non-perturbative,
resummation, and fixed-order QCD effects.
We observe good agreement
between the $b$-space resummation calculation using the published
values of the non-perturbative parameters from Ladinsky-Yuan
and the measurement for all values of \pt. Using their
parameterization for the non-perturbative region, we obtain
$g_2=0.59\pm0.06$ \gevsq.

\begin{table}
\begin{center}
\begin{tabular}{|c|c|c|c|}
   \pt\ range &  nominal \pt & number
  & ${d\sigma(Z\rightarrow e^+e^-)/dp_T}$ \\
   (\gevc) & value (\gevc) & of events
  & (pb/\gevc) \\
  \hline\hline
 0--1    &0.6& 156 & 6.04$\pm$0.53 \\
 1--2    &1.5& 424 & 16.2$\pm$0.96 \\
 2--3    &2.5& 559 & 20.4$\pm$1.1  \\
 3--4    &3.5& 572 & 19.7$\pm$1.1  \\
 4--5    &4.5& 501 & 16.2$\pm$0.92 \\
 5--6    &5.5& 473 & 15.0$\pm$0.87 \\
 6--7    &6.5& 440 & 14.1$\pm$0.84 \\
 7--8    &7.5& 346 & 11.1$\pm$0.73 \\
 8--9    &8.5& 312 & 10.0$\pm$0.69 \\
 9--10   &9.5& 285 & 9.29$\pm$0.67 \\
 10--12  &11.0& 439 & 7.25$\pm$0.54 \\
 12--14  &13.0& 326 & 5.45$\pm$0.44 \\
 14--16  &15.0& 258 & 4.45$\pm$0.39 \\
 16--18  &17.0& 203 & 3.54$\pm$0.33 \\
 18--20  &19.0& 181 & 3.21$\pm$0.31 \\
 20--25  &22.3& 287 & 2.06$\pm$0.18 \\
 25--30  &27.3& 174 & 1.29$\pm$0.13 \\
 30--35  &32.3& 124 & 0.962$\pm$0.11 \\
 35--40  &37.4& 104 & 0.840$\pm$0.10 \\
 40--50  &44.5&  92 & 0.373$\pm$0.045 \\
 50--60  &54.5&  61 & 0.251$\pm$0.036 \\
 60--70  &64.5&  40 & 0.163$\pm$0.027 \\
 70--85  &76.6&  20 & 0.053$\pm$0.012 \\
 85--100 &91.7&  13 & 0.034$\pm$0.009 \\
 100--200 &135& 15 & 0.0050$\pm$0.0013 \\
 200--300 &228&  2 & 0.0004$^{+0.0004}_{-0.0003}$ \\
\end{tabular}
\caption{Summary of the results of the measurement of the \pt\ distribution
of the \zb\ boson. The range of \pt\ corresponds to the intervals used for
binning the data. The nominal \pt\ corresponds to the value of \pt\ used
to plot the data and was obtained from theory.
The quantity ${d\sigma(Z\rightarrow e^+e^-)/dp_T}$
corresponds to the differential cross section in each bin of \pt\
for \zbee\ production. The uncertainty on the differential cross section
includes both systematic and statistical uncertatinties, but does not
include overall normalization uncertainty due to the luminosity of $\pm 4.4$\%.
 }
\label{tab:prl_results}
\end{center}
\end{table}

%
%
We thank the Fermilab and collaborating institution staffs for
contributions to this work and acknowledge support from the
Department of Energy and National Science Foundation (USA),
Commissariat  \` a L'Energie Atomique (France),
Ministry for Science and Technology and Ministry for Atomic
   Energy (Russia),
CAPES and CNPq (Brazil),
Departments of Atomic Energy and Science and Education (India),
Colciencias (Colombia),
CONACyT (Mexico),
Ministry of Education and KOSEF (Korea),
and CONICET and UBACyT (Argentina).

\end{document}